\pdfoutput=1

\documentclass[11pt]{article}

\usepackage[final]{acl}

\usepackage{times}
\usepackage{latexsym}

\usepackage[T1]{fontenc}

\usepackage[utf8]{inputenc}

\usepackage{microtype}

\usepackage{inconsolata}

\usepackage{graphicx}

\usepackage{subcaption}

\title{LLM-Assisted Translation of Legacy FORTRAN Codes to C++: A Cross-Platform Study}

\author{
  \textbf{Nishath Rajiv Ranasinghe\textsuperscript{1}},
  \textbf{Shawn M. Jones\textsuperscript{1}},
  \textbf{Michal Kucer\textsuperscript{1}},
  \textbf{Ayan Biswas\textsuperscript{1}}, \\
  \textbf{Daniel O'Malley\textsuperscript{1}},
  \textbf{Alexander Buschmann Most\textsuperscript{1}},
  \textbf{Selma Liliane Wanna\textsuperscript{1}},\\
   \textbf{Ajay Sreekumar\textsuperscript{2}}
\\
\textsuperscript{1}Los Alamos National Laboratory, Los Alamos NM 87545, \\
\textsuperscript{2}School of Information, University of Arizona, 103 E 2nd St \#4, Tucson, AZ 85721
}

\begin{document}
\maketitle
\begin{abstract} Large Language Models (LLMs) are increasingly being leveraged for generating and translating scientific computer codes by both domain-experts and non-domain experts. Fortran has served as one of the go to programming languages in legacy high-performance computing (HPC) for scientific discoveries. Despite growing adoption, LLM-based code translation of legacy code-bases has not been thoroughly assessed or quantified for its usability. Here, we studied the applicability of LLM-based translation of Fortran to C++ as a step towards building an agentic-workflow using open-weight LLMs on two different computational platforms. We statistically quantified the compilation accuracy of the translated C++ codes, measured the similarity of the LLM translated code to the human translated C++ code, and statistically quantified the output similarity of the Fortran to C++ translation.

\end{abstract}

\section{Introduction}
A Large volume of scientific computational software implemented in HPC environments has been written in programming languages such as Fortran and C due to their superior performance. However, recent advancements in computer hardware are not fully utilized by older generations of Fortran, and these legacy codes often encounter difficulties with memory allocations. There is a lack of human resources to maintain and improve these code-bases for mission critical applications in the future \cite{LANL_report, Fortran2Python}. 

Propriety (e.g. ChatGPT) and open weight (e.g. Llama \cite{Llama}) LLMs have vastly improved code generation \cite{code_generatation} and code translation between modern programming languages \cite{Jiao2023IsCA} due to widespread availability of training examples, but not without difficulties \cite{lost_in_translation}. As efforts expand to translate scientific software from legacy programming languages to more modern languages via agentic workflows, there is a need for systematic methods to evaluate the effectiveness of machine generated scientific software.

However, very few studies exist for LLM-assisted code translation from Fortran to C++, primarily due to a lack of quality training data sets. A recent study \cite{OpenMP_F2CPP}, compiled pairs of OpenMP Fortran and equivalent C++ codes to evaluate LLM code translation and evaluated their results using both quantitative (e.g., CodeBLEU score \cite{codebleu}) and qualitative approaches (e.g., human evaluation). There is also a lack of LLM-based Fortran to C++ code translation tools that can be readily deployed to assist developers in mission critical and secure environments. Furthermore, earlier attempts to translate code from Fortran to C++ have not accounted for successful compiles or output evaluation of the translated code \cite{F2C-painful}.

\begin{figure*}
    \centering
    \includegraphics[width=0.9\linewidth]{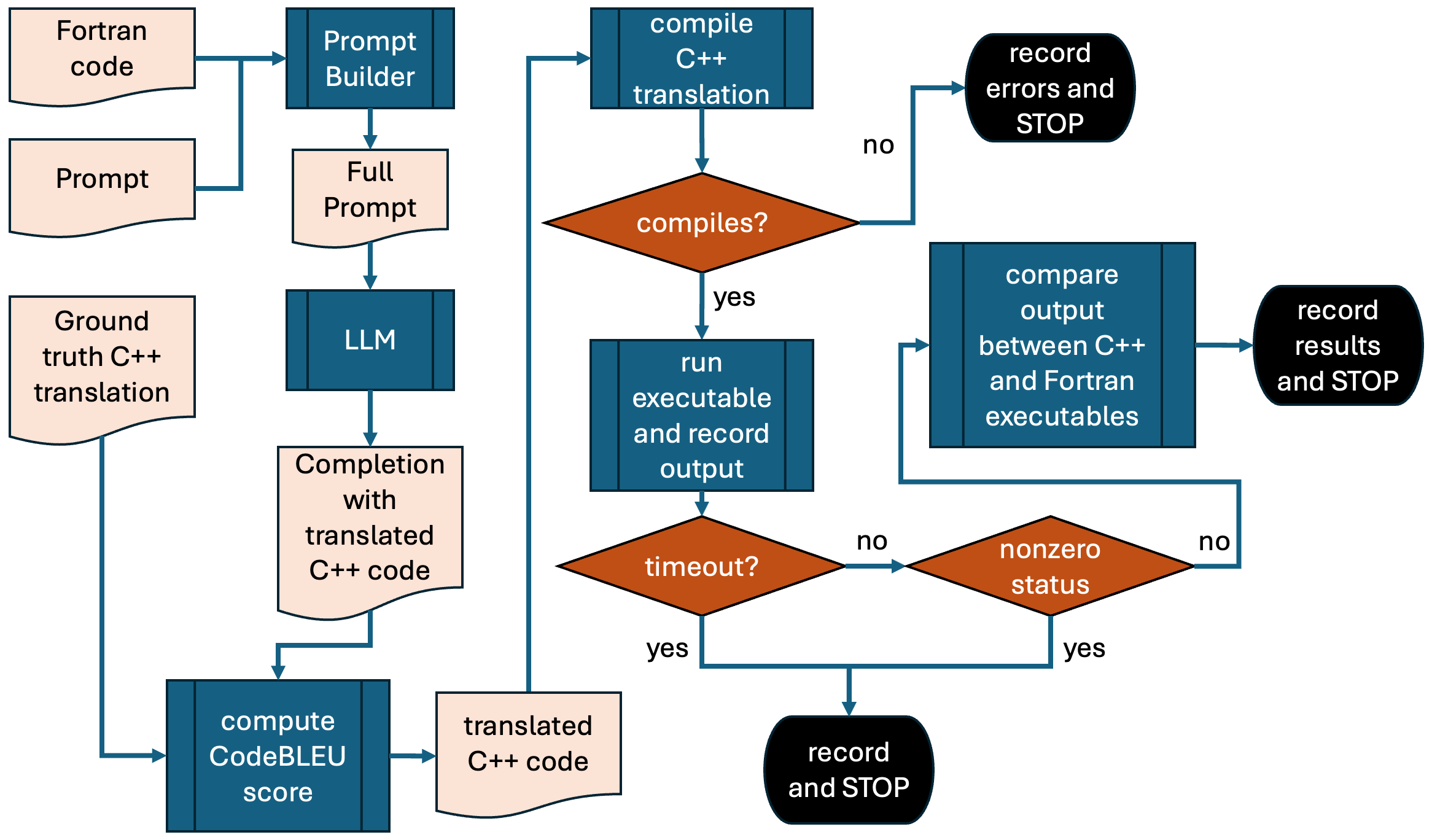}
    \caption{Regardless of LLM, our workflow evaluates several parts of the LLM's code translation, starting by comparing it to a human-translated ground truth with CodeBLEU, then moving to evaluate how well the translation compiles and executes. Finally, the workflow compares the output between the original Fortran code and the translated code's C++ executable.}
    \label{fig:flow-chart}
\end{figure*}

In this study, we make several contributions. We conduct an analysis of translating open-source code-bases using open-weight models.
Our workflow (Figure \ref{fig:flow-chart}) is designed to be agnostic of any specific LLM or computational platform (e.g., vLLM), building towards a set of standardized evaluation measures for machine-generated code translation. We evaluate the similarity to the human-translated target code using the common CodeBLEU measure \cite{codebleu}, how much of the translated code compiles (compilation accuracy \cite{BabelTowerLT}), and how well the output of the compiled translated code matches the original compiled Fortran code (output similarity). We also categorize any compile errors to demonstrate different behaviors among LLMs. To our knowledge, this is the first attempt to statistically quantify code translation accuracies of open-weight LLMs between computational platforms, the first such study involving Fortran, and the first to apply all of these evaluation techniques together.


\section{Background}


Despite the emergence of numerous modern programming languages, Fortran remains integral in legacy scientific applications, HPC, and areas requiring intensive numerical computations, such as climate modeling \cite{climate_FORTRAN}, computational fluid dynamics \cite{CFD_FORTRAN}, solving inverse problems \cite{cuer1980fortran}, full waveform inversion \cite{komatitsch2002spectral}, subsurface flow \cite{PFLOTRAN}, space applications \cite{copernicus},  crystallography \cite{Crystallography}, radiation transport \cite{mcnpx} and structural analysis \cite{nardelli1995parst95}. Unfortunately, Fortran is no longer a popular language \cite{LANL_report} and finding assistance from the community for future development is challenging. We chose C++ as a target language because it has more community support, but it also has a number of desirable features for scientific computing in the HPC environment, including its highly efficient feature set, template techniques \cite{C+_faster_than_FORTRAN}, the standard template library \cite{STL}, and advanced memory management \cite{memory_management_C++}. Unfortunately, efforts to translate legacy code-bases from Fortran to C++ have encountered several challenges stemming from differences in language paradigms, syntax, and standard libraries. 

LLMs have emerged as an efficient and robust method for translating code between programming languages. Many LLMs exist \cite{too_many_llms}, and there are different computational platforms \cite{emani2023_LLM} for executing LLMs. In this work, we evaluate two such platforms: vLLM and SambaNova. vLLM is a library providing a common interface for efficiently serving different LLMs across different hardware architectures utilizing the PagedAttention algorithm \cite{Vllm}. SambaNova is an AI accelerator platform that provides specialized hardware for executing LLMs \cite{Prabhakar_2024}. We compare both in this paper.

\section{Related Work}

Fortran to C++ translation has traditionally been conducted manually by experienced programmers. There have been few efforts to convert these legacy code-bases from Fortran to C++ using source-to-source translation tools \cite{FABLE, f2c}. However, the translated codes from these sources lack readability and require manual changes to implement memory management functionality \cite{F2C-painful}.

Previous systematic studies of code translation between pairs of modern programming languages such as C, C++, Go, Java, and Python using LLMs have been met with varying degree of compilation success from 2.1 to 47.3\% for code specific (codeGEN, CodeGenX, StarCoder) and text based general purpose (GPT-4, Llama-2, TB-Airboros, TB-Vicuna) LLMs with GPT-4 having the most success \cite{lost_in_translation}. Recent efforts to create larger code bases of example training data sets for popular and niche programming languages have improved the LLM assisted translations between more modern languages \cite{codetransocean}. A recent study \cite{fortran2cppautomatingfortrantocmigration} utilized an LLM based agentic method that seamlessly integrates multiple verification processes into iterative cycles for translating Fortran to C++. This approach employs a questioner-solver module to delegate referencing and decision-making tasks to separate LLMs, a multi-turn dialogue collection that effectively captures the nuanced aspects of translating and finally fine-tuning of three open-weight LLMs using the data produced to improve the accuracy of the models. Our study differs from theirs \cite{fortran2cppautomatingfortrantocmigration} by evaluating the capabilities of open-weight LLMs that can be readily deployed in a mission critical environment to translate Fortran to C++ on different computational architectures. We also differ by our choice in evaluations. We include compilation accuracy, the translated code's similarity to human translated codes, and a comparison of the similarity of outputs between our ground truth Fortran codes and the translated code from the LLM. Unlike other studies, we also apply the open-source Rosetta code repository \cite{RosettaCode_web} as a data source for evaluating the translation of Fortran to C++.

\begin{table*}[t]
\caption{The LLMs used in this study.}

  \label{tab:llms-in-this-study}
    \centering
\small
  \begin{tabular}{|l|l|l|}
    \hline
    \textbf{LLM} & \textbf{\# parameters} & \textbf{Computational platform} \\
    \hline \hline
    Open code interpreter     & 33B     & vLLM     \\  \hline
    Llama 3.1    & 70B    & vLLM    \\  \hline
    Mistral Large Instruct 2407      & 123B      & vLLM   \\  \hline
    Llama 3.3     & 70B    & vLLM     \\  \hline \hline
    Llama 3.1     & 8B     & SambaNova Cloud      \\  \hline
    Llama 3.1     & 70B    & SambaNova Cloud      \\  \hline
    Llama 3.1     & 405B   & SambaNova Cloud      \\  \hline
    Llama 3.3     & 70B    & SambaNova Cloud      \\  \hline
  \end{tabular}

\end{table*}

\begin{figure*}
  \centering
  \includegraphics[width=1.0\textwidth]{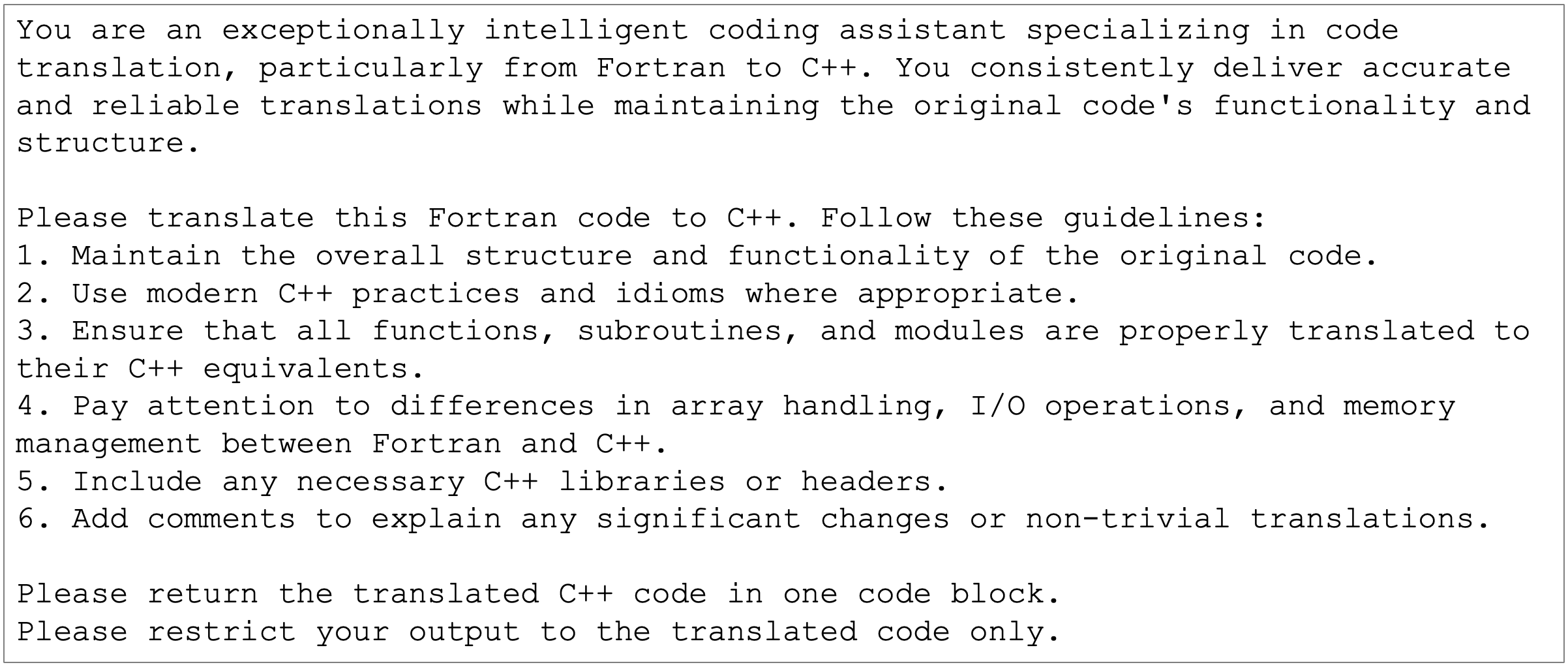}
  \caption{The prompt used in this study.}
  \label{fig:prompt}
\end{figure*}

\begin{table*}[t]
\caption{Classification of compiler errors used in this work.}
\label{tab:compiler-errors}
\scriptsize
\centering

\begin{tabular}{p{1in} p{2in} p{2.5in}}

\textbf{Compile Error Category} & \textbf{Error topic}         & \textbf{String matches from g++ compiler}            \\
\hline

Syntax Error                    & Missing operators, missing delimiters,  & expected\\
& incorrect usage of tokens,  & before\\
& or anything else resulting from poor programming syntax & error: no match for ‘operator\textgreater{}=\\
& & stray ‘`’ in program\\
& & error: void value not ignored as it ought to be\\
& & error: ‘std::std’ has not been declaredcannot be used as a function\\
& & error: assignment of read-only locationerror: invalid initialization of non-const reference of type\\
& & error: lvalue required as increment operand\\
& & error: no matching function for call to\\
& & error: missing terminating " character\\
& & error: too many arguments to function \\ \hline
Type Error                      & An issue with use of data types                                                                                           & invalid conversion \\
& & cannot convert \\ \hline
Linker Error                    & The implied use of external libraries & is not a member of ‘std’ \\
& & error: aggregate ‘std::stringstream ss’ has incomplete type and cannot be defined\\
& & undefined reference \\ \hline
Declaration Error               & Declaring variables before use                                                                                            & error: too many initializers\\
& & was not declared\\
& & has not been declared \\ \hline                                                                                                                   
Semantic Error                  & Proper application of functions or operators & invalid operands\\
& & invalid use of \\ \hline
Scope Error                     & Using variable outside of their established scope & not in this scope\\
& & is not captured \\ \hline                                                                                                                         
Template Error                  & Invalid use of C++ templates & wrong number of template arguments \\ \hline
File and I/O Error              & the code refers to nonexistent filesystem resources & No such file or directory\\ \hline
Memory Error                    & Incorrect use of memory operations & invalid use of\\
& & delete \\ \hline
Other Error                     & Anything else not covered with the string matching above & \\ \hline                                                                                                                                        
\end{tabular}

\end{table*}

\begin{figure*}[t]
    \centering
    \begin{subfigure}[t]{0.47\textwidth}
        \centering
        \vspace{0pt}
        \includegraphics[width=\textwidth]{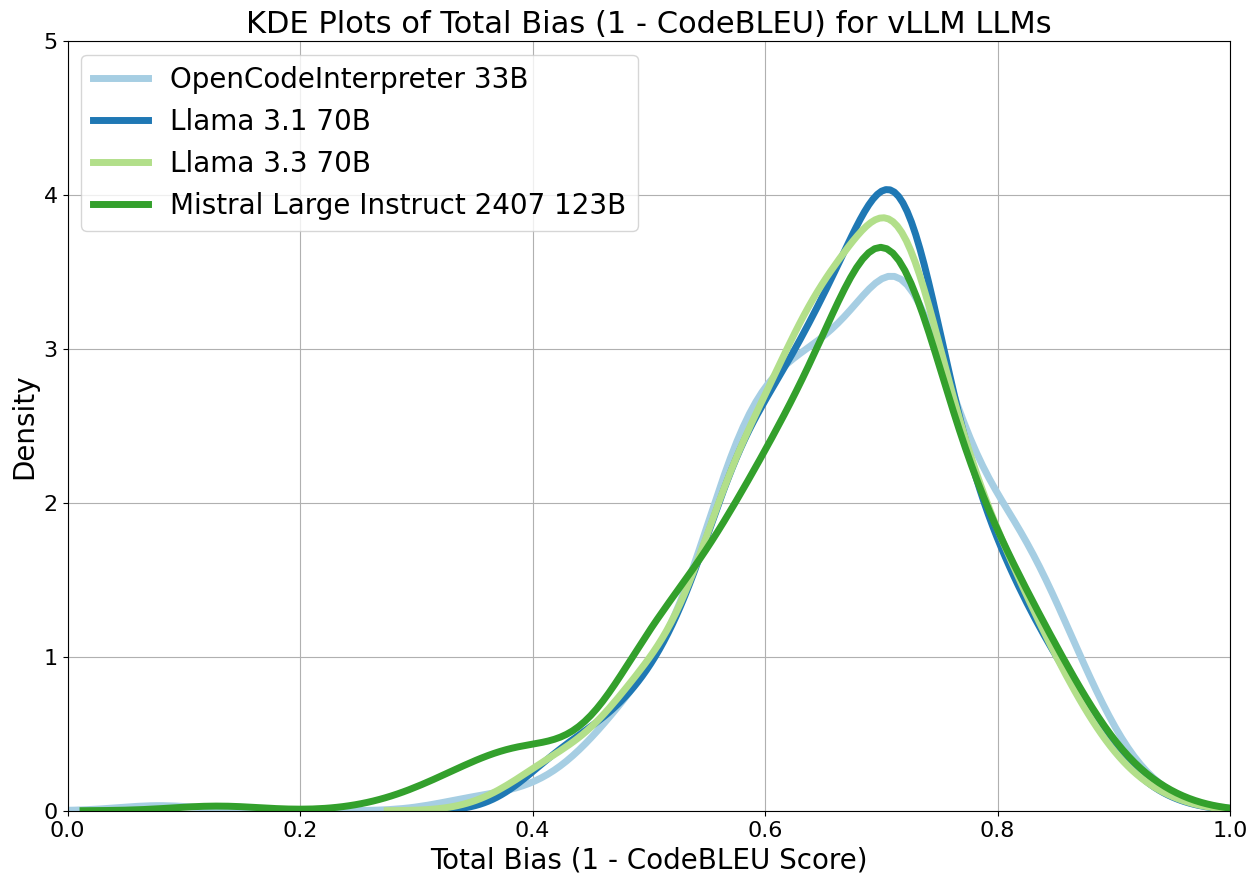}
        \caption{vLLM}
        \label{fig:kde-vllm}
    \end{subfigure}%
    \hfill
    \begin{subfigure}[t]{0.47\textwidth}
        \centering
        \vspace{0pt}
        \includegraphics[width=\textwidth]{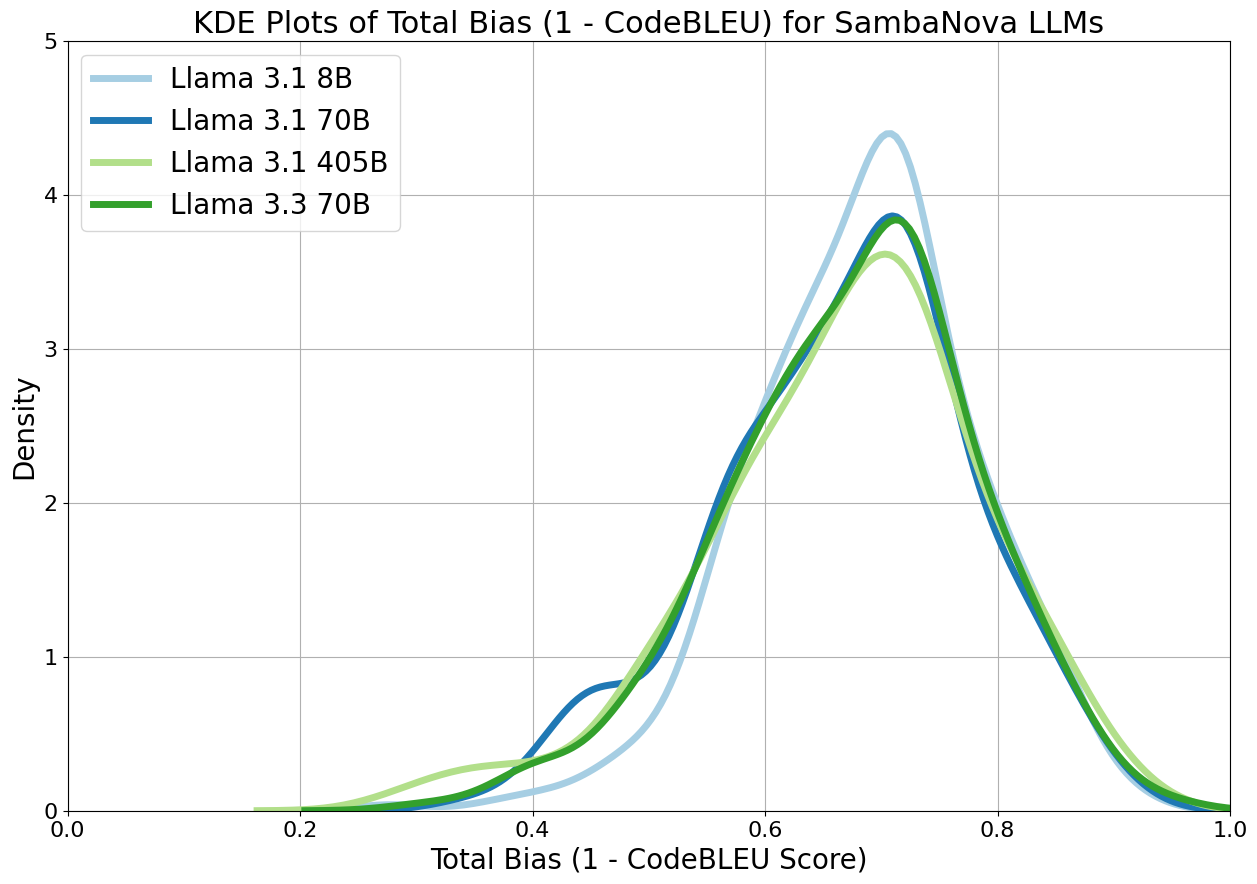}
        \caption{SambaNova}
        \label{fig:kde-sambanova}
    \end{subfigure}
    \caption{Kernel density estimate plots demonstrating the distribution of total bias (1 - CodeBLEU Score) for each Fortran translation demonstrates different distributions per execution platform.}
    \label{fig:kde-all}

    \centering
    \begin{subfigure}[t]{0.5\textwidth}
        \centering
        \vspace{0pt}
        \includegraphics[width=\textwidth]{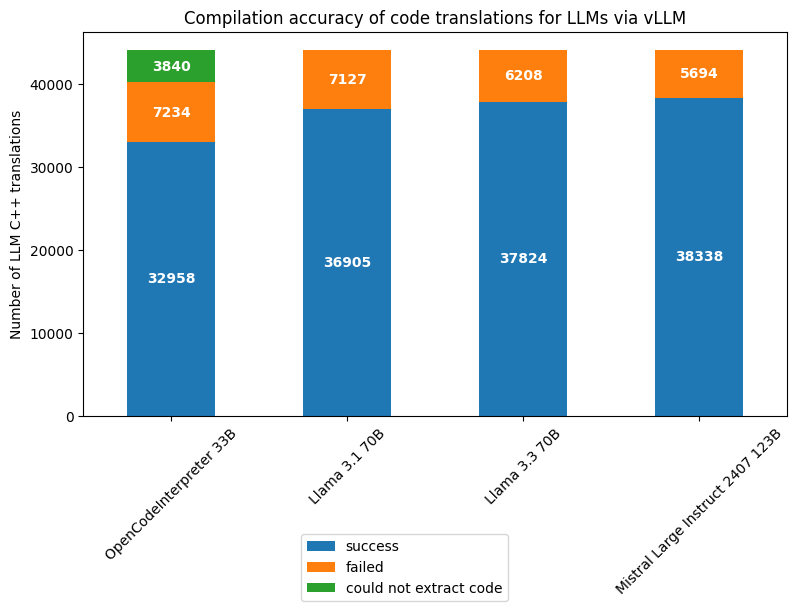}
        \caption{vLLM}
        \label{fig:compilation-accuracy-vllm}
    \end{subfigure}%
    ~ 
    \begin{subfigure}[t]{0.5\textwidth}
        \centering
        \vspace{0pt}
        \includegraphics[width=\textwidth]{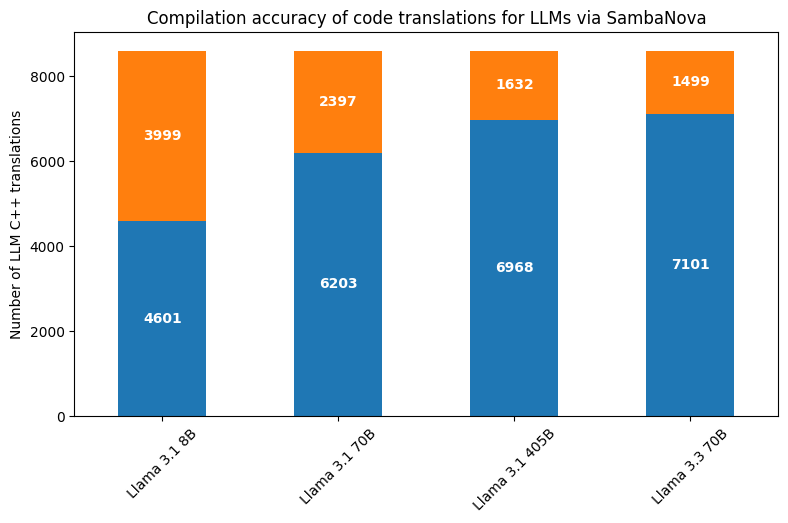}
        \vspace{1.2em}
        \caption{SambaNova}
        \label{fig:compilation-accuracy-sambanova}
    \end{subfigure}
    \caption{Compilation accuracy of each LLM by execution platform shows that the increase in the number of model parameters is proportional to the increase in compilation accuracy.}
    \label{fig:compilation-accuracy}
\end{figure*}

\begin{figure*}[t]
    \centering
    \begin{subfigure}[t]{0.5\textwidth}
        \centering
        \includegraphics[width=\textwidth]{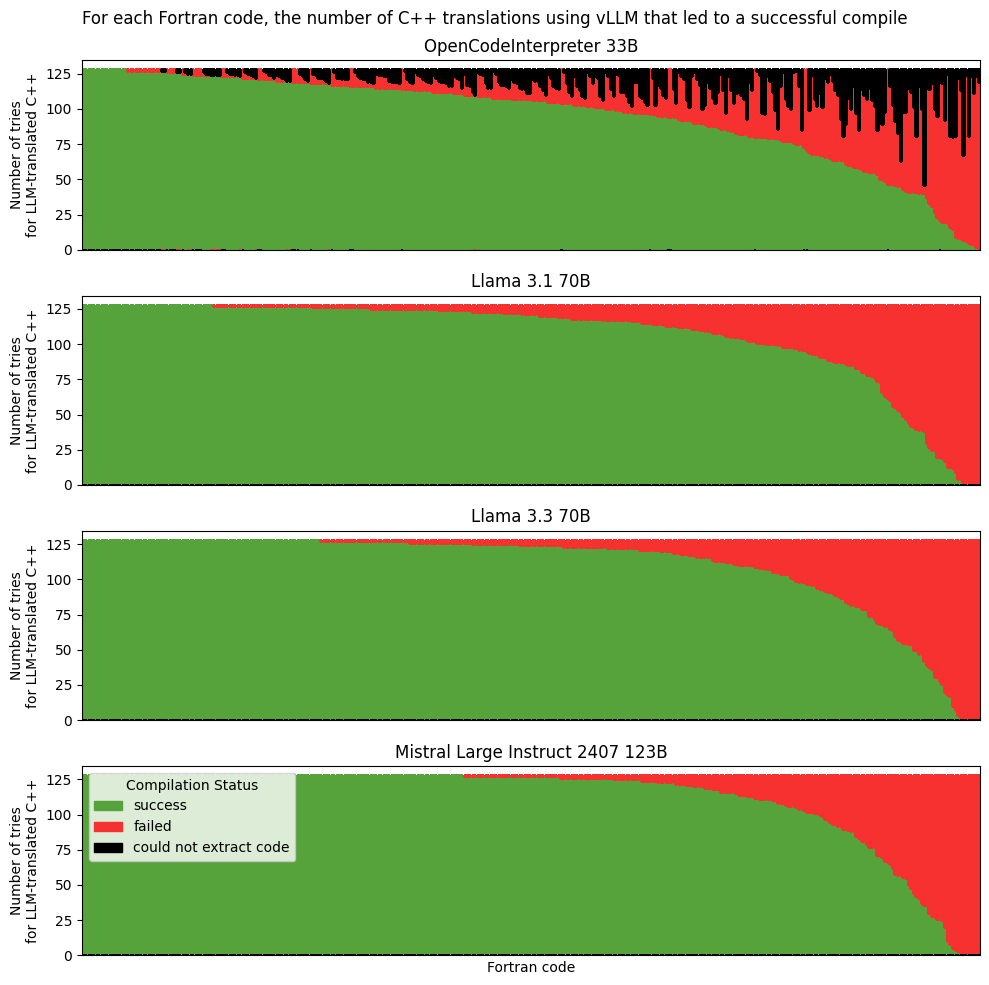}
        \caption{vLLM}
        \label{fig:compilation-accuracy-per-fortran-code-vllm}
    \end{subfigure}%
    ~ 
    \begin{subfigure}[t]{0.5\textwidth}
        \centering
        \includegraphics[width=\textwidth]{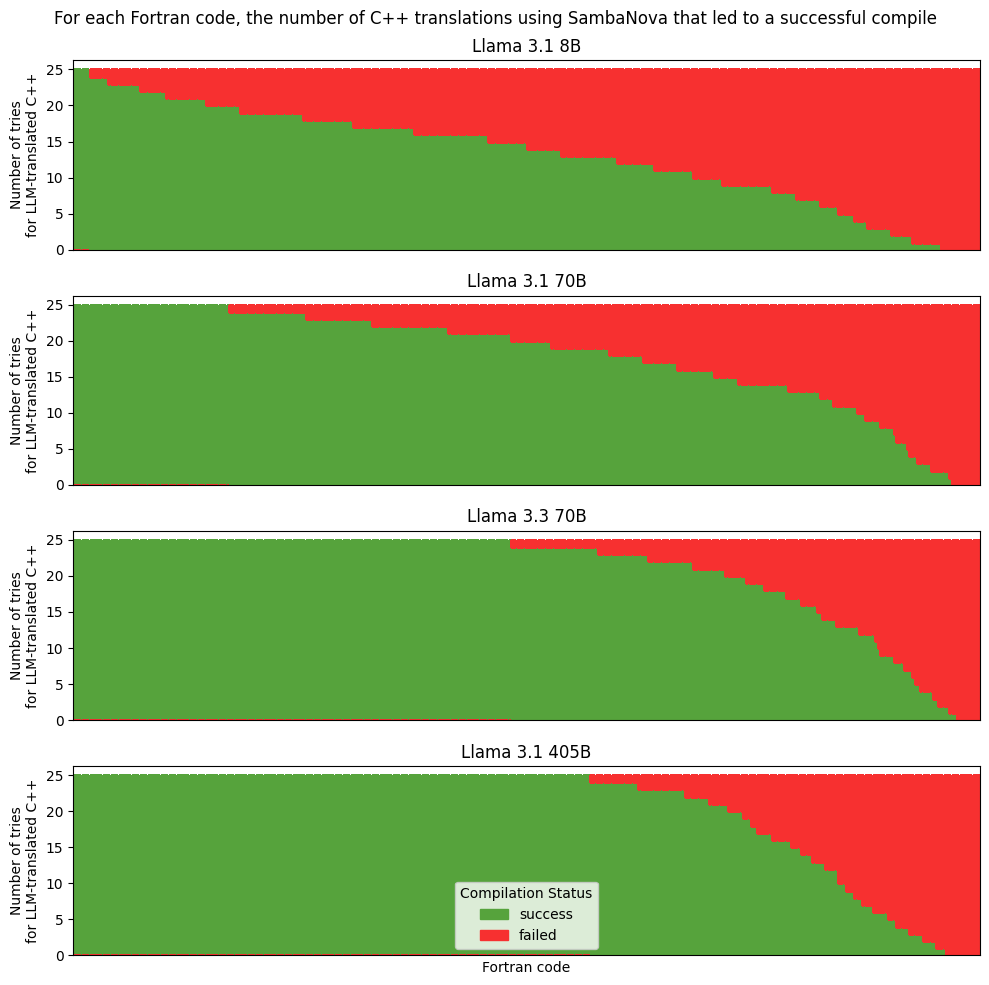}
        \caption{SambaNova}
        \label{fig:compilation-accuracy-per-fortran-code-sambanova}
    \end{subfigure}
    \caption{Each Fortran code is plotted along the x-axis while the count of tries for a  corresponding C++ translation is placed on the y-axis. Translations that compiled successfully are shown in green, and those that failed are marked in red. Note same Fortran code is not always shown at the same point in the x-axis. Compilation accuracy of each translated Fortran program differs per model with some LLMs having more difficulty translating certain codes than others. We note that LLMs with a higher number of parameters have more success per Fortran code.}
    \label{fig:compilation-accuracy-per-fortran-code}
\end{figure*}

\begin{figure*}[t]
    \centering
    \begin{subfigure}[t]{0.47\textwidth}
        \centering
        \vspace{0pt}
        \includegraphics[width=\textwidth]{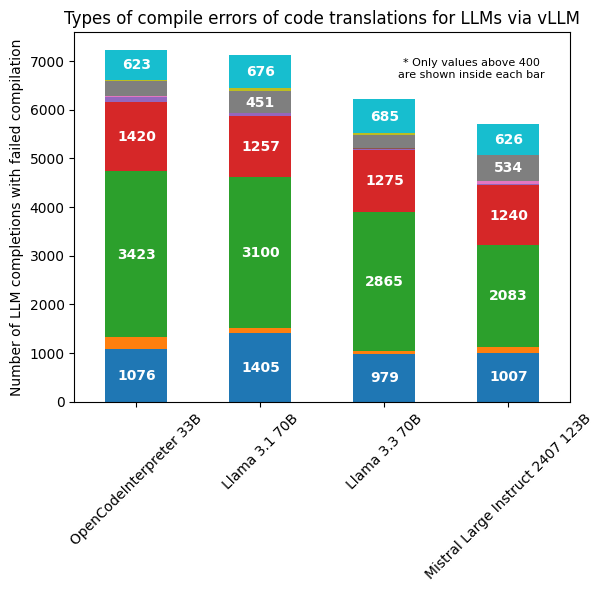}
        \caption{vLLM}
        \label{fig:compile-errors-vllm}
    \end{subfigure}%
    \hfill
    \begin{subfigure}[t]{0.5\textwidth}
        \centering
        \vspace{0pt}
        \includegraphics[width=\textwidth]{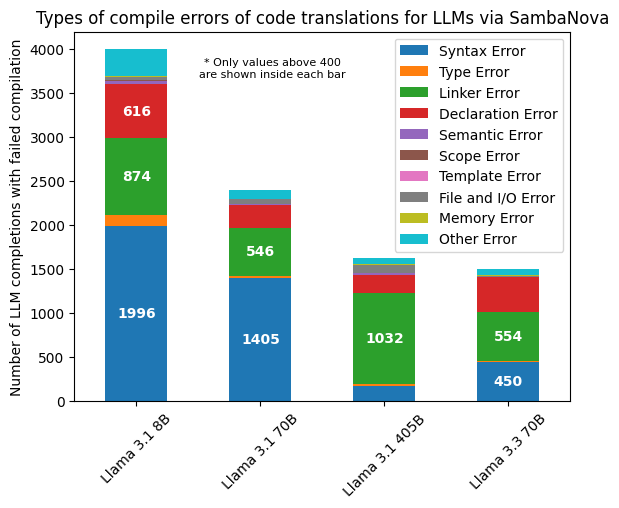}
        \vspace{1.2em}
        \caption{SambaNova}
        \label{fig:compile-errors-sambanova}
    \end{subfigure}
    \caption{Distribution of compile error categories for each C++ translation shows that LLMs produce different errors in their translated code.}
    \label{fig:compile-errors}
\end{figure*}

\begin{figure*}[t]
    \centering
    \begin{subfigure}[t]{0.47\textwidth}
        \centering
        \vspace{0pt}
        \includegraphics[width=\textwidth]{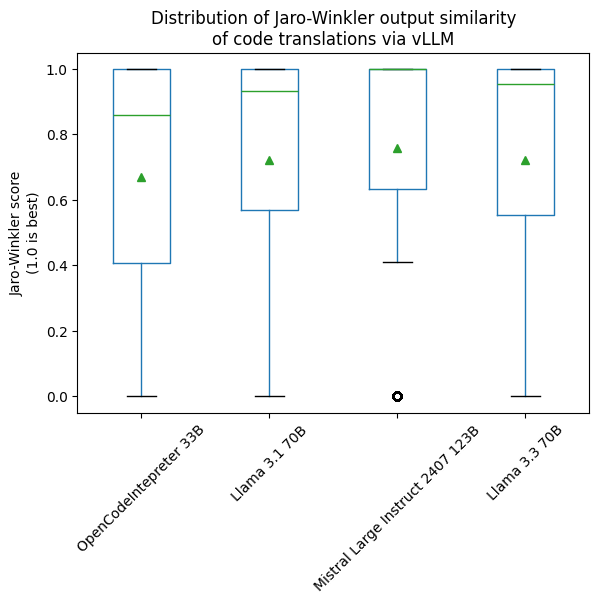}
        \caption{vLLM}
        \label{fig:output-similarity-distribution-vllm}
    \end{subfigure}%
    \hfill
    \begin{subfigure}[t]{0.47\textwidth}
        \centering
        \vspace{0pt}
        \includegraphics[width=\textwidth]{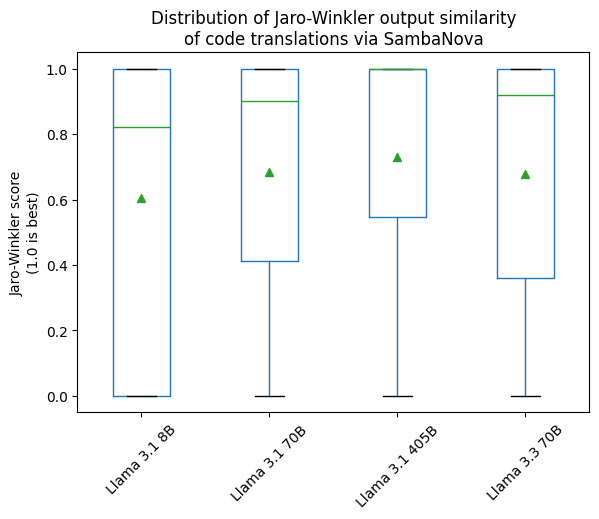}
        \vspace{1.5em}
        \caption{SambaNova}
        \label{fig:output-similarity-distribution-sambanova}
    \end{subfigure}
    \caption{Distribution of Jaro-Winkler scores for output similarity comparison between original Fortran executables and LLM C++ executables. Green triangles represent means while green lines are medians.}
    \label{fig:output-similarity-distributions}
\end{figure*}

\section{Methodology}


\subsection{Data}

To evaluate how well each LLM's translation matches a human translation, we required not only Fortran code, but ground truth C++ translations. We acquired two datasets containing pairs of Fortran and equivalent C++ codes. Rosetta Code \cite{RosettaCode_web} provides coding examples for the same programming task in multiple languages. We created a web scraper to produce a dataset of 243 Fortran and their corresponding C++ examples from the Rosetta Code website in October 2023. We retained only those examples for which there was at least one Fortran and corresponding C++ example per programming task. Our second dataset consists of 101 examples from the DataRaceBench (DRB) benchmark \cite{DRB} obtained from the OpenMP Fortran to C++ dataset \cite{OpenMP_F2CPP} that contains the same code implemented in different languages in support of the benchmark. From each dataset, we selected fully developed 344 computer programs with varying degrees of complexity, to ensure ground truth Fortran and C++ programs compile.

\subsection{LLMs}

Model parameters in LLMs are preset configurations that determine the model's architecture and training process, such as the number of layers, learning rate, and batch size. The number of parameters varies between LLMs. However, prior work \cite{Chinchilla} demonstrated that the performance of LLMs does not necessarily linearly increase with the number of parameters.

We chose LLMs that are well regarded by industry, can be deployed in a mission-critical environment, allow for local deployment to satisfy privacy concerns, have a diversity of model parameter sizes for comparison, and are also supported by the vLLM and SambaNova Cloud platforms \cite{sambanova_cloud}. Table \ref{tab:llms-in-this-study} shows the LLMs we selected based on this criteria. 




\subsection{Workflow}

Figure \ref{fig:flow-chart} shows the evaluation process we applied to each Fortran code and LLM. We start by building each full prompt by combining each Fortran code with the prompt in Figure \ref{fig:prompt}. Using this full prompt, we requested that each LLM convert the Fortran code to C++. Because LLMs are known to vary their responses due to their stochastic nature, we issued the same prompt multiple times for each Fortran code. We set up vLLM \cite{Vllm} using onsite hardware at the Los Alamos National Laboratory (DGX hardware equipped with 8 A100s NVIDIA GPUs along with 2 AMD EPYC 7742 64-Core Processors) and issued the same prompt 128 times per Fortran code per LLM. We utilized temperature of 0.8, min-p of 0.05, top-p of 0.95, and set the maximum generation length to 8192 tokens across the LLM models. We also used the OpenAI Python API library to prompt Llama models hosted by SambaNova Cloud, which is equipped with SambaNova SN40 Reconfigurable Dataflow Units (RDUs) \cite{Prabhakar_2024}. Due to rate limits on the SambaNova Cloud, we only executed the same prompt 25 times per Fortran code per LLM. We utilized temperature of 0.8, top-p of 0.9, and context length of 4096 across the Llama models in the SambaNova Cloud. From each completion, we recorded the C++ code and compared it to the ground truth C++ code from our datasets via CodeBLEU score \cite{codebleu}. From there, we evaluated the Fortran code's compilation accuracy and output similarity.


\subsection{Similarity to human translated code}

CodeBLEU \cite{codebleu} measures how well a machine translation matches a human translation for the same code. The CodeBLEU score contains four dimensions of comparison: matching n-grams, matching weighted n-grams, Abstract Syntax Tree matching, and data-flow analysis. We apply the human ground truth translation from each dataset to arrive at a CodeBLEU score. We perform bias analysis of the translated C++ codes across various LLMs, as an indicator of the code translation quality. We use CodeBLEU scores of the human translated C++ codes with their corresponding machine translated codes. In our scenario, since we run the same translation command prompt for a given code multiple times and we might get variations in the code translation, our bias analysis takes into account this stochasticity in LLM-based code generation. To perform this, for each LLM, we first calculate individual average CodeBLEU scores for each ground truth Fortran file across the trials. Since CodeBLEU depicts similarity, we calculate bias (that represents error) as $Bias = 1 - CodeBLEU$. With this formulation, now we can use these averaged bias scores to approximate a distribution using a non-parametric Kernel Density Estimate (KDE) approach\cite{KDE}. In this method, there exist different choices for its kernel types; such as Gaussian, triangular, rectangular, and the Epanechnikov kernel \cite{gramacki_KDE_book}. Generally, variations due to kernel types are considered to be less significant compared to the choice of kernel bandwidth \cite{silverman1998density}. Silverman’s rule of thumb for bandwidth selection generally produces smooth and good-quality density estimation \cite{biswas2016visualization}. We use this approach in our work and generate the KDE plots, as shown in Figure \ref{fig:kde-vllm} for vLLM based translated codes and Figure \ref{fig:kde-sambanova} shows the KDE plots for the SambaNova Cloud based translated codes. 


\subsection{Success of compilation}

Compilation accuracy of the translated C++ measures how many translations successfully compile without errors \cite{ParaBleu_for_Cuda}. We compiled each translated C++ using the g++ v5.3.0 compiler on Red Hat Enterprise Linux Workstation release 7.9. If a C++ translation failed to compile, we recorded the compiler output and did not proceed further with that translation (Figure \ref{fig:flow-chart}). We reviewed the compiler output and categorized each error as shown in Table \ref{tab:compiler-errors}. The 

\subsection{Similarity of outputs}

Output similarity compares the output of each Fortran program to that of its C++ translation generated from the LLM. We compiled each Fortran program and ran the resulting executable to capture its output. Then, we did the same with each LLM-generated C++ translation that successfully compiled. Outputs from scientific programs consist of text and numeric data. Humans may look at two outputs and consider them the same where a direct string match would score them radically different (e.g., \texttt{b(50,50)= 0.00000000} vs. \texttt{b(50,50)= 0.0} and \texttt{Fib for 30 832040} vs. \texttt{Fib for 30 = 832040.0}). We first tokenized each output using the NLTK \cite{NLTK} \texttt{word\_tokenize} function to produce a list of strings. Then, we attempted to convert each token to a floating point number using the Python \texttt{float} function. If the token could be converted, we rounded it to a precision of 4 decimal places. If not, then we left the token as a string. We, then applied a Jaro-Winkler \cite{jaro1989advances, winkler1990string} score to each set of tokens to measure their similarity.

Thus, by the end of the workflow we have evaluated each translation in comparison to a human translation, how well it compiles, and whether it produces the same output as the Fortran submitted to the system at the beginning.

\section{Results and Discussion}

\subsection{Similarity to human translated code}

CodeBLEU scores demonstrate how well an LLM's code translation matches a human translation of the same code. Figure \ref{fig:kde-all} shows the bias of CodeBLEU scores between LLMs. Scores on the x-axis provide a distance between LLM generated C++ translations and their human ground truth equivalents. Higher scores that indicate that the translation is farther than the ground truth and thus a poorer match. At first glance Figure \ref{fig:kde-all} appears to show that there is not much difference between LLMs, but the peaks give a more nuanced story.

Figure \ref{fig:kde-vllm} shows that Llama 3.1 70B leads with the highest rate of translations that do not match human ground truth. OpenCodeInterpreter 33B \cite{opencodeinterpreter} has the lowest peak outperforming Mistral Large. However, Mistral does have a small peak lower on the x-axis, indicating many more that might be closer to human ground truth.



SambaNova has a similar peak in Figure \ref{fig:kde-sambanova}, indicating a higher number of LLM translations that do not match human ground truth. Llama 3.1 8B's CodeBLEU bias is highest. Thus, its translations are least consistent with human translations. In contrast, Llama 3.1 405B has the lowest peak, but appears only marginally better in consistency than other models. These results with the commonly-used CodeBLEU metric demonstrate that larger models provide translations closer to human ground truth, but the amount of similarity in these distributions necessitate our other measures to more clearly separate performance.

\subsection{Success of compilation}

Figure \ref{fig:compilation-accuracy} shows the compilation accuracy results for each computational platform and LLM. In both cases, we see an increase in the number of successful compiles as one increases the number of parameters in the LLM. Additionally, as seen in Figure \ref{fig:compilation-accuracy-vllm}, while the LLMs served by vLLM appear to generate more successfully compilable code, OpenCodeInterpreter generates completions from which we cannot extract code. In contrast, SambaNova's results in Figure \ref{fig:compilation-accuracy-sambanova} show no instances where LLM completions produced code that could not be extracted. Additionally, we see that, for vLLM, Llama 3.1 70B and Llama 3.3 70B have comparable performance. This is not the case with these two LLMs on SambaNova Cloud, where Llama 3.1 405B and Llama 3.3 70B have similar performance.

Figure \ref{fig:compilation-accuracy-per-fortran-code} demonstrates the distribution of compilation accuracy for all Fortran codes. These sandcharts represent each Fortran code on the x-axis. The y-axis represents each translation of that code into C++. Green shows translations that successfully compile. Red shows failures. By executing each LLM multiple times we can see the level of variation in their responses and note that not all translation failures occurred equally. Some translations were always successfully compiled while others were more varied. We also note the same pattern of improving compilation accuracy among all Fortran codes as the number of parameters increases across models. vLLM shows more consistent translations (green rising closer to the top) while SambaNova shows a dramatic improvement for Llama 3.1 405B over Llama 3.3 70B that was not apparent in the raw numbers shown in Figure \ref{fig:compilation-accuracy-per-fortran-code-sambanova}.

Figure \ref{fig:compile-errors} shows the distribution and categorization of of compile failures. In Figure \ref{fig:compile-errors-vllm}, most of the compile errors generated from the LLMs served in vLLM are linker errors, representing the assumed inclusion of libraries not specified via an \texttt{\#include} directive. In contrast, in Figure \ref{fig:compile-errors-sambanova} the majority of the compile errors shown for LLMs served in SambaNova Cloud are syntax errors. Again, we see that Llama 3.3 70B and Llama 3.1 405B have comparable performance, though their compile error distribution varies. 

\subsection{Similarity of outputs}

Figure \ref{fig:output-similarity-distributions} shows the distribution of Jaro-Winkler scores comparing the outputs of the ground truth Fortran programs to the outputs of their LLM C++ translations. We note the same familiar pattern of increasing number of parameters leads to better mean similarity of inputs. Mistral Large with vLLM in Figure \ref{fig:output-similarity-distribution-vllm} and Llama 3.1 405B with SambaNova in Figure \ref{fig:output-similarity-distribution-sambanova} both outperform Llama 3.3 70B in this case. Mistral Large, however produces a tighter distribution of similar outputs.




\section{Conclusion}

We conducted an analysis of how well open-weight LLMs translate open-source code-bases from Fortran to C++. We presented an LLM-independent and platform-independent workflow for our evaluation. This workflow evaluates several elements of translation quality. We consider the similarity between human ground truth and machine translation, if the translated C++ code compiles, what errors are encountered if the compile fails, and finally how well the resulting C++ translation's executable produces the same output as the original Fortran code.

We ran this workflow with LLMs on both the vLLM and SambaNova Cloud platforms. Because LLMs do not always produce the same output each time, we ran 128 instances of the same translation on vLLM and 25 on SambaNova to ensure we had a sizeable sample space.  Unsurprisingly, we discovered that those LLMs with higher model parameter counts tend to produce better results. Our codeBLEU analysis reveals that Mistral Large served on vLLM and Llama 3.1 405B served on SambaNova Cloud produce codes that better matches human translations. Our compilation evaluation demonstrates that Mistral Large on vLLM and Llama 3.1 405B on SamaNova Cloud have higher counts of compilable code, with Llama-3.3 70B being comparable. We demonstrated that not all Fortran codes were translated consistently, showing that some LLMs produced C++ translations that more consistently compiled for a given Fortran code. We also found that the translated codes from vLLM that failed to compile mostly had linker errors while those from SambaNova largely contained syntax errors, even for the same LLM model. Finally, we showed that, for successful compiles, the output of the translated executables better matched the output of the original Fortran with Mistral Large on vLLM and Llama 3.1 405B on SambaNova Cloud, with Llama 3.3 70B being comparable on both platforms.

The implications for scientific computing are mixed. The state of the art shows the code bases in Fortran can be translated to C++ readily, but also demonstrate that no LLM on either platform was free of error. We still require a human-in-the-loop for code translation.




\section{Limitations} 
While our study presents a workflow for systematic evaluation of open-weight LLMs for Fortran-to-C++ code translation, there are several limitations that must be acknowledged: Our evaluation workflow is not yet packaged into a standalone tool that can provide Fortran-to-C++ translations along with compilation statistics and output similarity. Automating this workflow would make scientific discovery more accessible for researchers working in HPC environments. We did not present our attempts to improve compilation accuracy through agentic workflows by incorporating the error messages generated from compiling the codes produced by the LLM into a automatic dialog with the LLM. Our initial efforts in that direction were shown to increase the compilation accuracies of the translated codes and we are pursuing the agentic workflows in a future study.

Additionally, our study could be enhanced by incorporating more complex and extensive Fortran code-bases, such John Burkardt's data set \cite{burkardt_homepage} which are highly relevant to scientific computing. Furthermore, Chen et al. \cite{fortran2cppautomatingfortrantocmigration} showed that fine-tuning LLMs on Fortran to C++ datasets could improve each model's CodeBLEU scores by 1.5 to 3.3 times with up to a 92\% increase in successful compilations. Focusing our study's analysis on models which have been fine-tuned for Fortran to C++ translation could help create more useful tools for developers.

Further improvements could be made with prompt design and in this study, we used the same prompt for every LLM. It is possible that further exploration of prompt design could uncover that different models perform better with different prompts \cite{prompt, knobloch2025}. Our study focused solely on open-weight LLMs such as Llama and Mistral. While comparisons do exist for both natural language translation as well as coding (without translating), our literature review found a lack of studies comparing open-weight LLMs to proprietary models like GPT and Gemini for code translation. Including these models, along with the source-to-source translation tools \cite{f2c, FABLE} which were popular for Fortran to C++ in the past could provide a clearer benchmark for our results. Additionally, in this study, we did not test the capabilities of the new generation of reasoning models (OpenAI’s o1, o1-mini, o3-mini; DeepSeek-R1; and Anthropic Claude 3.7 Sonnet) to translate Fortran to C++. However, our workflow delivers a plug-and-play solution to test any LLMs code translation capabilities on any computational platform without any modifications. 

In this study, we did not consider improving code translation accuracy using few-shot learning via Retrieval Augmented Generation (RAG) as it is studied elsewhere \cite{Manish_paper}.

\section{Acknowledgments}
This work was supported by the Computational Systems and Software Environments subprogram of National Nuclear Security Administration's (NNSA's) Advanced Simulation and Computing program through Los Alamos National Laboratory (LANL). LANL is operated by Triad National Security, LLC, for the National Nuclear Security Administration of U.S. Department of Energy (Contract No. 89233218CNA000001). This research used resources provided by the Darwin testbed and DGX pod at LANL which is funded by the Computational Systems and Software Environments subprogram of LANL’s Advanced Simulation and Computing program (NNSA/DOE). We are also grateful to SambaNova Systems, Inc for providing access to SambaNova Cloud and technical support. This work is approved for unlimited release with an LA-UR number LA-UR-25-22376.

\newpage
\bibliography{custom}

\appendix



\end{document}